\documentclass{article}
\usepackage{natbib,epsfig,multirow}

\usepackage{amsfonts}
\usepackage{natbib}
\usepackage{colonequals}
\usepackage{amsfonts} 
\usepackage{amsmath}
\usepackage{subfigure}
\usepackage{amssymb,lineno}
\usepackage[usenames,dvipsnames]{xcolor}
\usepackage{mathrsfs}
\usepackage{booktabs}
\usepackage{graphicx}
\usepackage{subfig}
\usepackage{multirow}
\usepackage{array}
\newcommand{\matsig}{\mathbf\Sigma}

\def\EM{{EM}}

\newcommand{\bs}{\boldsymbol}
\newcommand{\skw}{\bs{\lambda}}

\newcommand{\Y}{\bs{Y}}
\newcommand{\X}{\bs{X}}

\newcommand{\y}{\textbf{\textit{y}}}

\newcommand{\z}{\textbf{\textit{z}}}

\newcommand{\Tr}[1]{\text{tr}\Big\{ #1 \Big\}}

\newcommand{\te}{$t$EIGEN} 
\newcommand{\MCLUST}{MCLUST} 
\newcommand{\stclust}{S$t$CLUST} 
\newcommand{\snclust}{SNCLUST} 
\newcommand{\skewtCLUST}{S$t$CLUST} 
\newcommand{\skewt}{skew-$t$}
\newcommand{\Skewt}{Skew-$t$}
\newcount\colveccount
\newcommand*\colvec[1]{
        \global\colveccount#1
        \begin{pmatrix}
        \colvecnext
}

\newcommand{\sbrac}[1]{\big[ #1 \big]}

\def\EM{{EM}}
\newcommand{\bic}{{BIC}}
\newcommand{\m}{{M}}

\newcommand{\vecmu}{\mbox{\boldmath$\mu$}}

\newcommand{\vecx}{\mathbf{x}}

\newcommand{\vecz}{\mathbf{z}}

\newcommand{\varthet}{\mbox{\boldmath$\vartheta$}}

\newcommand{\matd}{\mathbf{D}}
\newcommand{\mata}{\mathbf{A}}

\begin{document}

\title{Parsimonious Skew Mixture Models for\\ Model-Based Clustering and Classification}
\author{Irene Vrbik and Paul~D.~McNicholas\thanks{E-mail: paul.mcnicholas@uoguelph.ca. Tel: +1-519-824-4120, ext.\ 53136.}}
\date{Department of Mathematics \& Statistics, University of Guelph.}

\maketitle

\begin{abstract}
In recent work, robust mixture modelling approaches using skewed distributions have been explored to accommodate asymmetric data.  We introduce parsimony by developing \skewt\ and skew-normal analogues of the popular GPCM family that employ an eigenvalue decomposition of {a positive-semidefinite} matrix. The methods developed in this paper are compared to existing models in both an unsupervised and semi-supervised classification framework. Parameter estimation is carried out using the expectation-maximization algorithm and models are selected using the Bayesian information criterion. The efficacy of these extensions is illustrated on simulated and benchmark clustering data sets.
\end{abstract}

\section{Introduction}
\label{sec:1}
The objective of cluster analysis is to organize data into groups wherein the similarity within groups and the dissimilarity between groups are maximized.  A `model-based' approach is one that uses mixture models for clustering. The use of finite mixture models has become increasingly common for clustering and classification, particularly with the use of Gaussian components. The Gaussian model-based clustering likelihood is 
\begin{equation}\label{eqn:gmm}
\mathcal{L}(\varthet\mid\vecx_1,\ldots,\vecx_n)=\prod_{j=1}^n\sum_{i=1}^g\pi_i\phi(\vecx_j\mid\vecmu_i,\matsig_i),
\end{equation}
where  $\pi_i >0$, such that $\sum_{i=1}^g \pi_i = 1$, are mixing proportions and $\phi(\vecx_j\mid\vecmu_i,\matsig_i)$ is the density of a multivariate Gaussian random variable with mean~$\vecmu_i$ and covariance matrix $\matsig_i$. 
Model-based classification is a semi-supervised version of model-based clustering (cf.\ Section~\ref{sec:mbclass}).

Gaussian mixture models have been used for a wide variety of clustering applications, including work by \cite{mclachlan1988}, \cite{bouveyron07}, \cite{mcnicholas2008parsimonious,mcnicholas2010model,mcnicholas10d}, and \cite{baek2010}, amongst others. In efforts to accommodate data that exhibit some departure from normality, robust extensions are garnering increased attention. For instance, mixtures of multivariate $t$-distributions \citep{mclachlan98,peel} have proven effective for dealing with components containing outliers.  They have been the basis of a variety of robust clustering techniques that use mixtures of multivariate $t$-distributions, including work by \cite{mclachlan2007}, \cite{andrews2011FA,andrews11b}, \cite{baek11}, \cite{steane12}, and \cite{mcnicholas2012}, amongst others. 

Capturing components that are asymmetric can be tackled using skew-normal distributions \cite[cf.][]{lin2007} or other non-elliptically contoured distributions \citep[e.g.,][]{karlis09}. 
One example of a non-elliptical distribution is the skew-normal independent distribution, as considered for finite mixture modeling in \cite{cabral2012multivariate}. Another interesting alternative is the skew Student-$t$-normal distribution that has recently been used to model skewed heavy-tailed data \citep*{ho2012maximum}. Although there are many non-symmetric options available, we focus on mixtures of skew-normal distributions and mixtures of \skewt\ distributions herein. 

Recently, mixtures of multivariate \skewt\ distributions have been receiving some attention in the literature. Of course, the skew-normal, $t$, and Gaussian distributions are all special cases of the \skewt\ distribution.  This property can be important in clustering applications because we often do not know the most appropriate underlying distribution.  Alongside the skewness parameter that accommodates asymmetric data, the degrees of freedom parameter allows for heavy tails, giving less weight to outlying observations in parameter estimation. Compared to work on parsimonious Gaussian and $t$-mixtures, the literature contains relatively little on parsimonious mixtures of multivariate skew-normal and skew-$t$ distributions. The purpose of this paper is to go some way towards addressing this deficiency.
  
The remainder of the paper is organized as follows.  In Section~\ref{skewt}, we introduce \skewt\ and skew-normal mixture models and briefly discuss the calculation of parameter estimates.  Section~\ref{methodology} presents the construction of parsimonious families of models that are analogues of popular Gaussian approaches. The proposed methods are compared with Gaussian and multivariate $t$ analogues using simulation studies (Section~\ref{simulation}) and four benchmark clustering data sets (Section~\ref{applications}).  These models are extended further to semi-supervised classification in Section~\ref{sec:mbclass}, which is followed by concluding remarks (Section~\ref{conclusions}).

\section{Mixtures of Skew-$t$ and Skew-Normal Distributions}
\label{skewt}

 \subsection{A Mixture of \Skewt\ Distributions}\label{skewtdist}
 
 Although the computational tractability of Gaussian mixture models has contributed to their widespread popularity within the literature, their application is not always appropriate. For instance, \cite{kotz} argue that the multivariate-$t$ distribution provides a more realistic model for real-world data and it has been noted \cite[e.g.,][]{lin2007} that Gaussian mixture models have a tendency to over fit skewed data. Therefore, it is natural to consider a single distribution, namely the multivariate \skewt\ distribution, that conflates the robust properties of the $t$-distribution with a skewness parameter to account for asymmetry. We outline a model-based approach using parsimonious mixtures of multivariate \skewt\ distributions; we also consider a mixture of skew-normal distributions, which is a limiting case.

There are a number of `skew-$t$' distributions within the literature. The version that we adopt, as defined by \cite{pyne2009}, uses a particular stochastic representation of the multivariate \skewt\ distribution given by \cite{sahu}. This distribution also corresponds to the so-called `restricted' multivariate skew-$t$ distribution, as defined in  \cite{lee2012finite}.  Adopting this characterization, a random vector $\Y$ is said to follow a $p$-variate \skewt\ distribution with location vector $\bs{\xi}$, scale matrix ${\bs{\Omega}}$, skewness vector $\skw$, and $\nu$ degrees of freedom if it has the representation
\begin{equation}
\Y = \skw|U| + \X,
\label{eq:heir}
\end{equation}
where 
\[
  \begin{pmatrix}
  \X\\
 U\\
 \end{pmatrix}
\sim \mathcal{N}\left(  
\begin{pmatrix}
 \boldsymbol{\xi}\\
 0\\
 \end{pmatrix},
 \begin{pmatrix}
  \boldsymbol{\bs{\Omega}} & \textbf{0} \\
 0 & 1
 \end{pmatrix} \frac{1}{w}\right),
\]
and $W \sim \Gamma(\nu/2, \nu/2)$. 

We consider a $g$-component mixture of $p$-dimensional skew-$t$ distributions with density given by
\begin{equation}
f(\boldsymbol{y}_j\mid \bs{\Psi}) = \sum_{i = 1}^g \pi_i\varkappa(\boldsymbol{y}_j\mid\bs{\xi}_i,{\bs{\Omega}}_i,{\skw}_i,\nu_i),
\end{equation}
where $\varkappa(\boldsymbol{y}_j\mid\bs{\xi}_i,{\bs{\Omega}}_i,{\skw}_i,\nu_i)$ is the density of a multivariate skew-$t$ distribution with location vector $\bs\xi_i$, {scale}  matrix $\bs\Omega_i$, skewness parameter $\bs\lambda_i$, and $\nu_i$ degrees of freedom, and $\bs\Psi$ contains all model parameters, i.e., $\bs\Psi$ contains the parameters $\{\pi_i,\bs\xi_i,\bs\Omega_i,\bs\lambda_i,\nu_i : i=1,\ldots,g\}.$

\subsection{A Mixture of Skew-Normal Distributions}
Now consider the skew-normal distribution, which is a limiting case of the skew-$t$ distribution. This version of the skew-normal distribution is the `restricted' version of the skew-normal distribution defined by \cite{sahu} and proposed for the analysis of flow cytometric data by \cite{pyne2009}.  Resembling the above characterization (Section~\ref{skewtdist}), a random vector $\Y$ is said to follow a $p$-variate skew-normal distribution with location vector $\bs{\xi}$, {scale} matrix ${\bs{\Omega}}$, and skewness vector $\skw$ if it has the representation
\begin{equation}
\Y = \skw|U| + \X,
\label{eq:heirsn}
\end{equation}
where
\[
  \begin{pmatrix}
  \X\\
 U\\
 \end{pmatrix}
\sim \mathcal{N}\left(  
\begin{pmatrix}
 \boldsymbol{\xi}\\
 0\\
 \end{pmatrix},
 \begin{pmatrix}
  \boldsymbol{\bs{\Omega}} & \textbf{0} \\
 0 & 1
 \end{pmatrix}\right),
\]

We consider a $g$-component mixture of $p$-dimensional skew-normal distributions. The density is given by
\begin{equation}
f(\boldsymbol{y}_j\mid \bs{\Theta}) = \sum_{i = 1}^g \pi_i\varphi(\boldsymbol{y}_j\mid\bs{\xi}_i,{\bs{\Omega}}_i,{\skw}_i),
\end{equation}
where $\varphi(\boldsymbol{y}_j\mid\bs{\xi}_i,{\bs{\Omega}}_i,{\skw}_i,\nu_i)$ is the density of a multivariate skew-normal distribution with location vector $\bs\xi_i$, {scale} matrix $\bs\Omega_i$, and skewness $\bs\lambda_i$, and $\bs\Theta$ contains all model parameters, i.e., $\bs\Theta$ contains $\{\pi_i,\bs\xi_i,\bs\Omega_i,\bs\lambda_i : i=1,\ldots,g\}.$

\subsection{Parameter Estimation}

Parameter estimation for mixtures of multivariate skew-normal distributions is carried out via an expectation-maximization (\EM) algorithm \citep{dempster77} as outlined by \cite{pyne2009}. However, maximum likelihood estimation for mixtures of multivariate skew-$t$ distributions is more involved and has been tackled using a number of different variations of the \EM\ algorithm. \cite{vrbik2012} derive closed form solutions for the skew-$t$ random variable defined in (\ref{eq:heir}), allowing for a traditional \EM\ algorithm to be used. This aforementioned procedure is employed for our proposed study but multiple techniques are available. For example, a Bayesian approach is explored in \cite{fruhwirth2010bayesian}, wherein an efficient Markov chain Monte Carlo (MCMC) scheme is proposed for finite mixtures of multivariate skew-normal and skew-$t$ distributions.

A nice overview of the differing characterizations of the skew-$t$ distribution and their applications has recently been given by \cite{lee2012finite}. Apart from providing an up-to-date account of the recent work with the skew-$t$ distribution, they also develop some new results on the restricted and unrestricted multivariate skew-$t$ distribution (rMST and MST, respectively). The {E}-step for mixtures of MST can be expressed in closed form apart from one term and, when using a one-step-late M-step, this term can be taken to be zero or be expressed in closed form. \cite{lee2012finite} offer an alternative method to calculate the intractable E-step for the rMST.  This is accomplished by manipulating the expectations to be expressed in terms of the moments of the truncated non-central multivariate $t$-distribution \cite*[cf.][]{ho}.  

\subsection{Comments}
\begin{figure}[htbp]
\begin{center}
\includegraphics[width=0.8\linewidth]{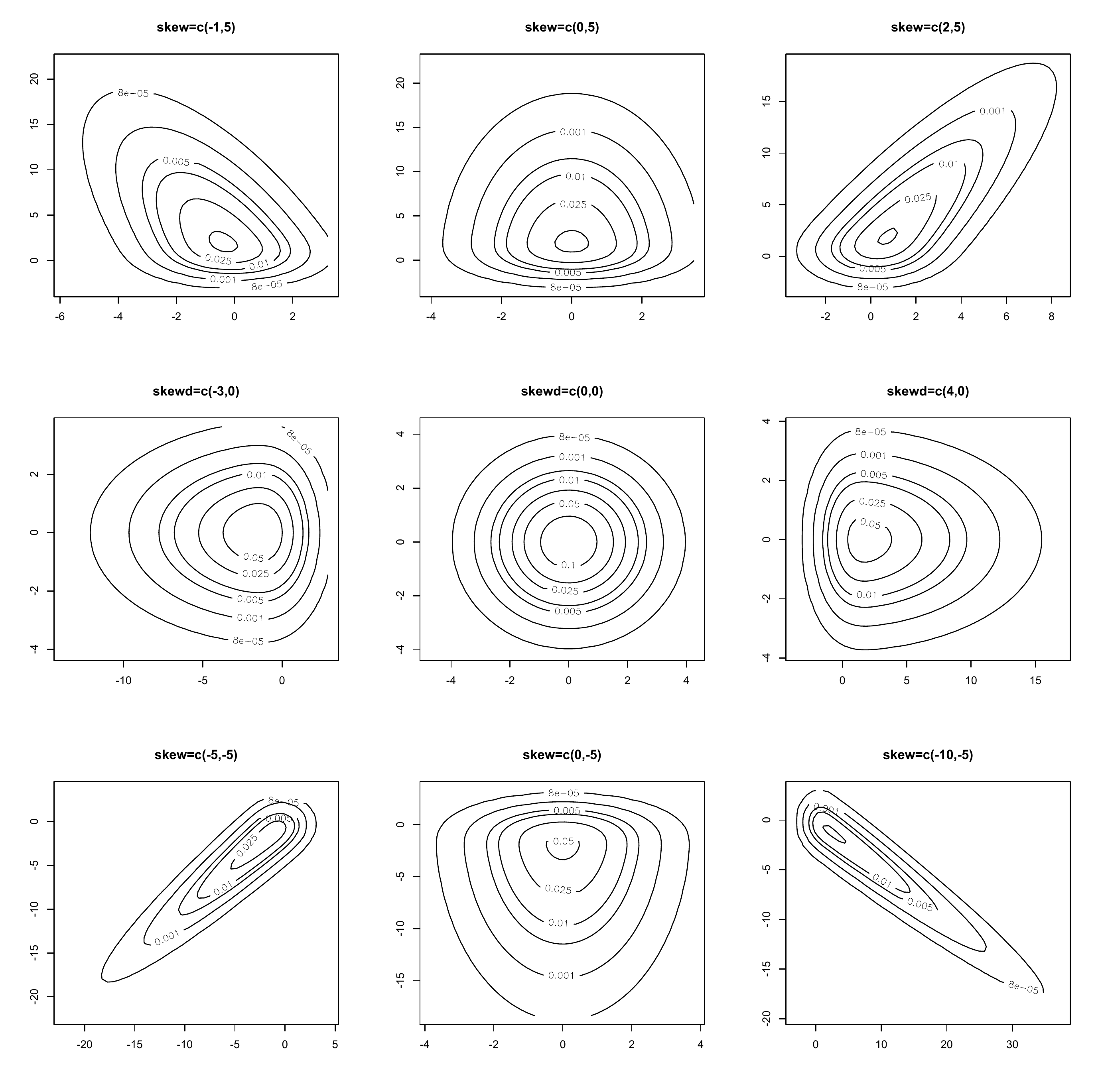}
\caption{Density contours from bivariate skew-normal distributions with $\bs\xi = \bs0$, $\bs\Omega=\mathbf{I}$, and varying values for the skewness parameter ($\skw$).}
\label{contours}
\end{center}
\end{figure}

The additional skewness parameter gives these models the advantage of being able to adjust for asymmetric data while the degrees of freedom in the \skewt\ case acts as a robustness tuning parameter to accommodate heavier tails. The contour maps in Figure \ref{contours} illustrate various shapes attained by the bivariate skew-normal distribution, with teardrop-like as well as spherical and elliptically shaped clusters.  The pictures are constructed using a multivariate skew-normal distribution with $\bs\Omega = \mathbf{I}$ to isolate the effect of skewness on otherwise elliptically shaped clusters. Note that a multivariate skew-$t$ distribution with large degrees of freedom would appear similar, but we would see more in the tails with lower values of degrees of freedom.

\section{Methodology}
\label{methodology}

\subsection{Two Skewed Families}
Despite the fact that there has been support for model-based approaches for quite some time now, only in the last two or more decades has this approach become popular. 
Owing greatly to the advances in computing, model-based approaches are quickly becoming an attractive tool for clustering and classification.  For Gaussian model-based clustering, model fitting becomes an issue due to the number of parameters in the component covariance matrices: there are $Gp(p+1)/2$ in all, cf.\ (\ref{eqn:gmm}). This same problem manifests for mixtures of multivariate $t$, skew-normal, and skew $t$-distributions. 

\cite{celeuxclust} introduced parsimony into Gaussian mixtures by imposing constraints on eigen-decomposed component covariance matrices. 
This parameterization, originally considered in \cite{Banfield1993}, is an eigenvalue decomposition of the component covariance matrices given by $\bs{\Sigma}_i = \lambda_i\matd_i\mata_i\matd_i'$, where $\matd_i$ is the orthogonal matrix of eigenvectors of $\bs{\Sigma}_i$, $\mata_i$ is the diagonal matrix of entries proportional to eigenvalues with $|\mata_i|=1$, and $\lambda_i$ is the associated constant of proportionality. \cite{celeuxclust} developed a family of fourteen Gaussian parsimonious clustering models (GPCMs) by applying different constraints to this eigen-decomposed covariance structure (Table~\ref{ta:steigen}).  A subset of these models make up the of family ten models available in the {\tt mclust} package \citep{fraley2012mclust} for {\sf R} \citep{R11}. In addition to clustering applications, Gaussian mixture models with this covariance decomposition have been used for discriminant analysis \citep{Bensmail1996} and classification \citep{dean06}. 
\begin{table}[htb]
\caption{The nomenclature used in the MCLUST family and borrowed for the models developed herein, along with the {decomposition constraints} and number of free parameters in the decomposed matrix. Note: `E' denotes equal across groups, `V' denotes varying across groups, `I' denotes the identity matrix.}
\label{ta:steigen}
{\scriptsize\begin{tabular*}{1.0\textwidth}{@{\extracolsep{\fill}}lcccl}\hline
Model & $\lambda_i=\lambda$ & $\mata_i = \mata$ & $\matd_i=\matd$ &   Free Covariance/Scale Parameters\\\hline
EII & E & I & I &  $1 $\\
VII & V & I & I &  $G$\\
EEI & E & E & I  & $p$\\
VEI & V & E & I  & $G + (p-1)$\\
EVI & E & V & I & $ Gp - (G-1)$\\
VVI & V & V & I &  $Gp $\\
EEE & E & E & E  & $\ \ [p(p+1)/2]$\\
VEE* & V & E & E &$\ \ [p(p+1)/2]+(G-1)$ \\
EVE$*^+$ & E & V & E & $\ \ [p(p+1)/2]-(G-1)(p-1)$\\ 
VVE$*^+$ & V & V& E & $\ \ [p(p+1)/2]+(G-1)p$ \\
EEV & E & E & V  & $G[p(p+1)/2]-(G-1)(p)$\\
VEV & V & E & V  & $G[p(p+1)/2]-(G-1)(p-1)$\\
EVV* & E & V& V& $G[p(p+1)/2]-(G-1)$\\
VVV & V & V & V  & $G[p(p+1)/2]$\\
\hline
\end{tabular*}}
{\scriptsize *These models are not considered in \MCLUST. \qquad\qquad
+These models are not considered in \te.}
\end{table}

\cite{andrews11c} introduced the $t$EIGEN family, which comprises $t$-analogues of the MCLUST models, along with two other GPCM models, with the added restriction of constraining or not constraining the component degrees of freedom $\nu_i$ to be equal across groups.  The $t$EIGEN family is supported by the{\tt teigen} package \citep{teigen} for {\sf R}.
The nomenclature for the $t$EIGEN models consist of four letters: `C'	denotes that a constraint is imposed, `U' denotes that a constraint is not imposed, and `I' denotes that the matrix in question is taken to be the identity matrix of suitable dimension. For each member of the $t$EIGEN family, relevant letters indicate the constraints on $\lambda_i, \matd_i, \mata_i$, and $\nu_i$, respectively. 

Herein, we consider parsimonious families of multivariate skew-normal and skew-$t$ distributions that are non-elliptical, robust extensions to the GPCM family; we refer to our families as the \snclust\ and \stclust\ families, respectively.
Note that the geometric interpretation of the component shapes for members of the \snclust\ and \stclust\ families is not the same as for members of the GPCM, MCLUST, and $t$EIGEN families unless the skewness is zero. 
In the GPCM family, the constraints (cf.\ Table~\ref{ta:steigen}) are applied to the decomposed component covariance matrices and so can be considered in terms of the volume, shape, and orientation of the component densities. However, for the \snclust\ and \stclust\ families, we are decomposing the component scale matrices ${\bs{\Omega}}_i$; accordingly, the shapes of the corresponding component densities cannot be interpreted after the fashion of MCLUST (or GPCM) unless the skewness is zero \cite[cf.][]{wang2009multivariate}. This lack of geometric interpretability is of no importance because the rationale for imposing constraints on the decomposed component scale matrices herein is simply to introduce parsimony. 

\subsection{Parameter Estimation}

This section outlines the parameter estimation for the \stclust\ family.  Estimates for the \snclust\ family can be derived in a similar fashion. First, note that, from the representation given in (\ref{eq:heir}), the joint distribution of $\y,u,w$ can be written
\begin{equation*}\begin{split}
f({\y}, u,w)  &=  f({\y}| u, w)f( u|w)f(w)\\
&= \frac{|w|^{1/2}}{(2\pi)^{p/2}|\bs{\Omega}|^{1/2}}\exp\left\{ -\frac{1}{2} ({\y} -{\bs\xi} -{\skw}|u|)' \left(\frac{\bs{\Omega}}{w}\right)^{-1}({\y} -{\bs\xi}- {\skw}|u|)   \right\} \\
   &\qquad\qquad\qquad\qquad\qquad\qquad\qquad\times \frac{1}{\sqrt{2\pi(1/w)}}\exp\left\{-\frac{u^2}{2(1/w)}\right\}\frac{\left(\frac{\nu}{2}\right)^{{\nu}/{2}}}{\Gamma\left(\frac{\nu}{2}\right)}w^{\frac{\nu}{2}-1}\exp\left\{-\frac{\nu}{2} w\right\}\notag\\
 &= \frac{(\frac{\nu}{2})^{{\nu}/{2}}w^{({\nu+p-1})/{2}}}{(2\pi)^{\frac{p+1}{2}}|{\bs\Omega}|^{\frac{1}{2}}\Gamma\left(\frac{\nu}{2}\right)}\exp\left\{ -\frac{w}{2} \big[({\y} -{\bs\xi} -{\skw}|u|)'{\bs\Omega}^{-1}({\y} -{\bs\xi} -{\skw}|u|)  + u^2 + \nu\big] \right\}.
\end{split}\end{equation*}

Within the \EM\ framework, it is convenient to adopt the following notation. We regard the observed data $\y = (\y_1',\ldots,\y_n')$ as being incomplete, where $\bs u = (u_1,\ldots,u_n)$ and $\bs w= (w_1,\ldots,w_n)$ are unobservable latent variables and the component membership labels $\z = (\z_1,\ldots,\z_n)$ are the missing data. Note that $\z$ is defined such that $\z_j = (z_{1j},\ldots, z_{gj})'$ for $j = 1,\ldots,n$, where $z_{ij}=1$ if $\y_j$ belongs to component~$i$ and $z_{ij}=0$ otherwise. 
Now, $(\y,\bs u,\bs w,\z)$ is referred to as the complete-data and the complete-data log-likelihood for unconstrained component scale matrices $\bs\Omega_1,\ldots,\bs\Omega_g$ is given by
\begin{equation}\begin{split}
\log \mathscr{L}(\bs{\Psi}\mid \y,\bs u,\bs w,\z) &= \sum_{i=1}^{g}\sum_{j=1}^{n} z_{ij} \log \pi_i + \sum_{i=1}^{g}\sum_{j=1}^{n} z_{ij} \log f_i(\boldsymbol{y}_j, u_j, w_j)=\mathscr{L}_{1} + \mathscr{L}_{2} + \mathscr{L}_{3},
\label{log} 
\end{split}\end{equation}
say, where
\begin{align*}
\mathscr{L}_{1}
=&  \sum_{i=1}^{g} \sum_{j=1}^{n} z_{ij}\log\pi_i,\\
\mathscr{L}_{2}
=& \sum_{i=1}^{g} \sum_{j=1}^{n} z_{ij}\Big\{-\frac{1}{2}\Big[\log(2\pi) +  \log |{\bs \Omega_i}^{-1}| + w_j({\y}_j - {\bs\xi}_i - {\skw}_iu_j)'{\bs\Omega_i}^{-1} ({\y}_j - {\bs\xi}_i - {\skw}_iu_j) \Big]\Big\},\\
\mathscr{L}_{3}
=&  \sum_{i=1}^{g} \sum_{j=1}^{n} z_{ij}\Big\{-\frac{1}{2}\sbrac{(p-1)\log(w_j) + w_iu_j^2}
+ \frac{\nu_i}{2}\sbrac{w_j - \log(\nu_i/2)} + \log\Gamma(\nu_i/2)+ (\nu_i/2 - 1)\log(w_j)\Big\}. 
\end{align*}

In the {E}-step, the expected value of the complete-data log-likelihood can be found using any of the three methods mentioned in Section~\ref{skewt}; we use the method of \cite{vrbik2012} in our analyses (Sections~\ref{simulation} and~\ref{applications}). The updates for $\pi_i$, $\bs\xi_i$, $\skw_i$, and $\nu_i$ needed in the \m-step are given by \cite{wang2009multivariate}; parameter estimation for the decomposed elements of $\bs\Omega_i=\lambda_i\matd_i\mata_i\matd_i'$ will depend on the model under consideration (cf.\ Table~\ref{ta:steigen}) and can be found in an analogous fashion to the Gaussian case. 
To acquire the updates for $\lambda_i$, $\matd_i$, and $\mata_i$, we need only maximize $\mathscr{L}_2$ in (\ref{log}), which is equivalent to minimizing
\begin{align*}
\sum_{i=1}^{g}\sum_{j=1}^{n}  z_{ij} \log |{\bs \Omega_i}^{-1}| +  &\sum_{i=1}^{g} \sum_{j=1}^{n}  z_{ij}w_j({\y}_j - {\bs\xi}_i - {\skw}_iu_j)'{\bs\Omega_i}^{-1} ({\y}_j - {\bs\xi}_i - {\skw}_iu_j)\\
=&\sum_{i=1}^{g} \sum_{j=1}^{n}  z_{ij} \log |{\bs \Omega_i}^{-1}| +  \Tr{\sum_{i=1}^{g} \sum_{j=1}^{n}  z_{ij} w_j({\y}_j - {\bs\xi}_i - {\skw}_iu_j)'{\bs\Omega_i}^{-1} ({\y}_j - {\bs\xi}_i - {\skw}_iu_j)}\\
=&\sum_{i=1}^{g} \sum_{j=1}^{n}  z_{ij} \log |{\bs \Omega_i}^{-1}| +  \Tr{\sum_{i=1}^{g}  W_i {\bs\Omega_i}^{-1} },
\end{align*}
where $W_i = \sum_{j=1}^{n} z_{ij} w_j({\y}_j - {\bs\xi}_i - {\skw}_iu_j)({\y}_j - {\bs\xi}_i - {\skw}_iu_j)'$.  Finally, define $W \colonequals\sum_{i=1}^{g}W_i$ 
and note that parameter estimation for $\lambda_i$, $\matd_i$, and $\mata_i$ becomes identical to that carried out by \cite{celeuxclust}.  

\subsection{Model Selection}
The best model from amongst the ten members of the \stclust\ family is determined by the popular Bayesian information criterion \citep[\bic;][]{schwarz}.   Although the underlying regularity conditions for the asymptotic approximation are not generally satisfied
\cite[cf.][]{keribin00}, the BIC has shown to be a useful tool for selecting amongst mixture models \citep[e.g.,][]{fraley02a,andrews2011FA}.
 The BIC is given by $-2l(\vecx, \hat{\varthet})+m\log n$,
where $m$ is the number of free parameters, 
$l(\vecx,\hat{\varthet})$ is the maximized log-likelihood, and $\hat{\varthet}$ is the maximum likelihood estimate of $\varthet$. \cite{dasgupta98} propose using the BIC for mixture model selection. When defined as above, the model with the largest BIC is selected. 
Model selection for the \snclust\ family is handled in the same way.
 
\subsection{Performance Assessment}
Because the true classes for the data sets used in the applications that follow are known, the adjusted Rand index \citep[ARI;][]{hubert85} is used to assess the clustering results. The ARI is a popular measure of classification agreement between the true and predicted group memberships or, more generally, between any two partitions. The ARI is a corrected form of the Rand index \citep{rand71} that adjusts for chance agreement. An ARI of 1 corresponds to perfect agreement whereas an ARI of 0 corresponds to results no better than would be expected by guessing.
 
\section{Simulation Studies}
\label{simulation}

\subsection{Introduction}
Although the proposed methods can explicitly account for skewed components, they should also be capable of capturing symmetric components, e.g., data from a multivariate Gaussian or $t$-distribution.  In Section~\ref{sim1}, we illustrate that the \stclust\ and \snclust\ families can uncover underlying Gaussian and $t$-mixtures. We then demonstrate the difficulties encountered by the MCLUST and $t$EIGEN families when faced with skewed data (Section~\ref{sim2}). The {\tt mclust} and {\tt teigen} packages are used to run the MCLUST and $t$EIGEN models, respectively. In Section~\ref{sim3}, we present a more difficult data set in which we compare the four algorithms.
Note that to facilitate direct comparison with MCLUST, we restrict the \te, \snclust, and \stclust\ families to ten models to correspond to the \MCLUST covariance structures (cf.\ Table~\ref{ta:steigen}) for all our simulations.

\subsection{Simulation 1}
\label{sim1}
For this simulation, the first component, of size $n_1$ = 300, was generated from a bivariate normal distribution with parameters 
$$\vecmu_1=\begin{pmatrix} 3\\ 0 \end{pmatrix}, \quad\matsig_1 = \begin{pmatrix} 1&0\\ 0&1 \end{pmatrix},$$ 
and the second component, of size $n_2=200$, was generated from a bivariate $t$-distribution with $\nu=4$ degrees of freedom and parameters 
$$\vecmu_2=\begin{pmatrix} -2\\ 4 \end{pmatrix}, \quad \matsig_2=\begin{pmatrix} 1&0.5\\ 0.5&1 \end{pmatrix}.$$
 
\begin{figure}[htbp]
\begin{center}
\includegraphics[width=0.75\linewidth]{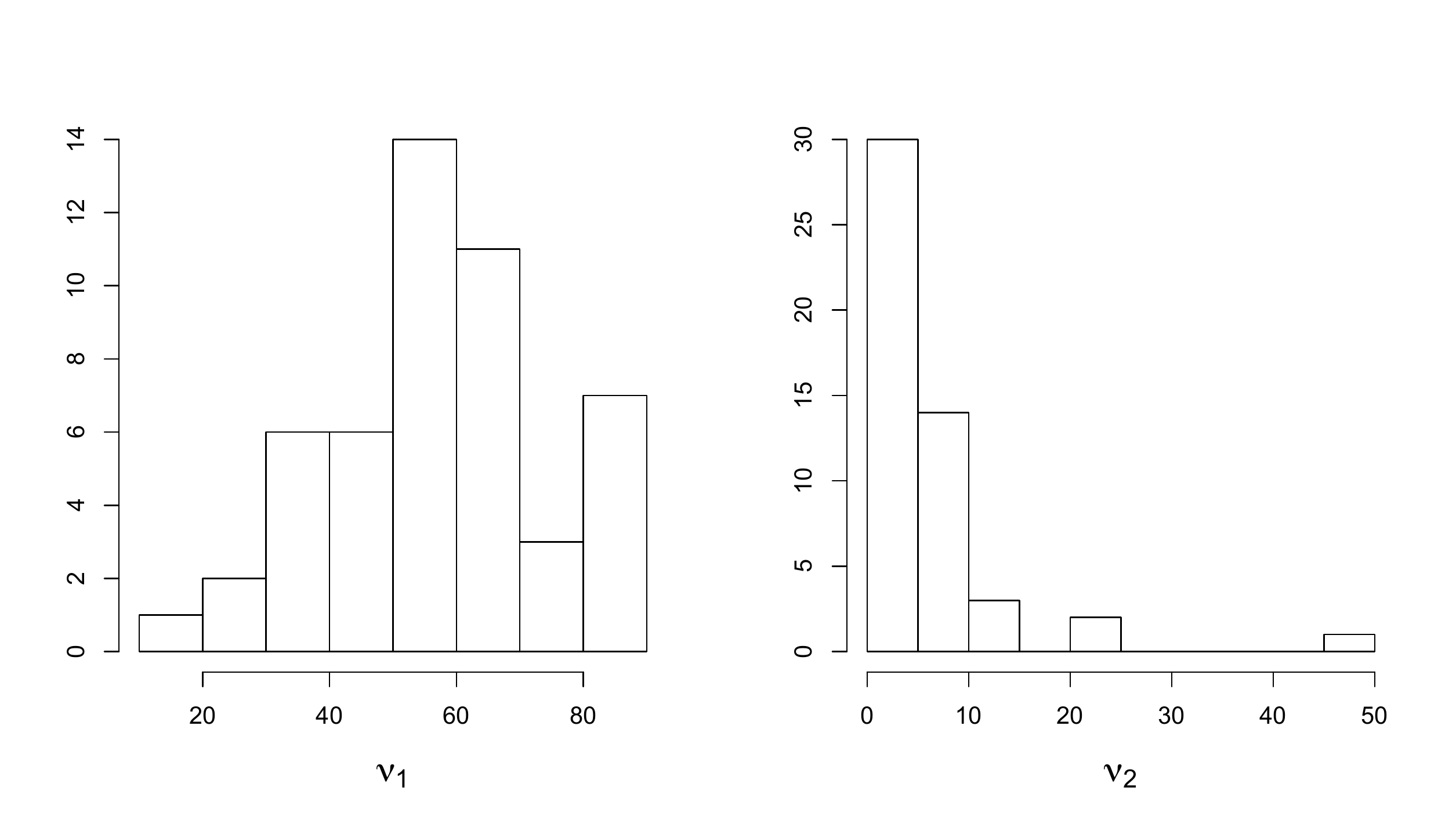}
\vspace{-0.1in}
\caption{Histogram of the estimates for the degrees of freedom parameter from fitting the \stclust\ models on the 50 random runs for Simulation~1.} 
\label{histdf}
\end{center}
\end{figure}

The clustering performance of the \stclust\ family was excellent, yielding an average ARI of 0.99 with standard deviation 0.007. The average fitted values for degrees of freedom were $\hat\nu_1 = 56.92$ and $\hat\nu_2 =  6.51$ and, looking across all runs, sensible values were almost always selected (Figure~\ref{histdf}). The average values for skewness were $\hat\skw_1 = (0.021, -0.035)$ and $\hat\skw_2 = (0.014, 0.011)$, which makes sense when fitting elliptical components.  
\vspace{-0.2in}

The clustering performance of the \snclust\ family was also very good despite the fact that one of the components was simulated from a multivariate $t$-distribution.

\subsection{Simulation 2}
\label{sim2}
In this section, skewed data were generated from bivariate skew-$t$ distributions using the convenient representation given in (\ref{eq:heir}). In all, we used 50 random simulations with three components of sizes $n_1 = 150$, $n_2 = 200$, and $n_3 =150$, respectively, with locations $\bs\xi_1 = (0,0), \bs\xi_2 = (6,25), \bs\xi_3 = (4,0)$, skewness vectors $\bs\skw_1 = (2,4), \bs\skw_2 = (-2,4), \bs\skw_3 = (2,4)$, degrees of freedom $\nu_1=10, \nu_2 = 8, \nu_3 = 70$, and scale matrices 
\begin{equation*}
\bs\Omega_1= \left( \begin{array}{ccc}
0.4 & 0.2\\
0.2 & 0.5\end{array} \right),\qquad
\bs\Omega_2= \left( \begin{array}{ccc}
1 & 0.5\\
0.5 & 1\end{array} \right),\qquad
\bs\Omega_3= \left( \begin{array}{ccc}
0.2 & 0\\
0 & 0.3\end{array} \right).
\end{equation*}
Note that the third group has high degrees of freedom and is effectively a skew-normal cluster. All families were run for $i = 1,\dots,9$ components; as a matter of practice, when the BIC chose a $g=9$ component model, the algorithm was rerun for $i=1,\dots,12$. 

\begin{figure}[htbp]
\begin{center}
\includegraphics[width=1\linewidth]{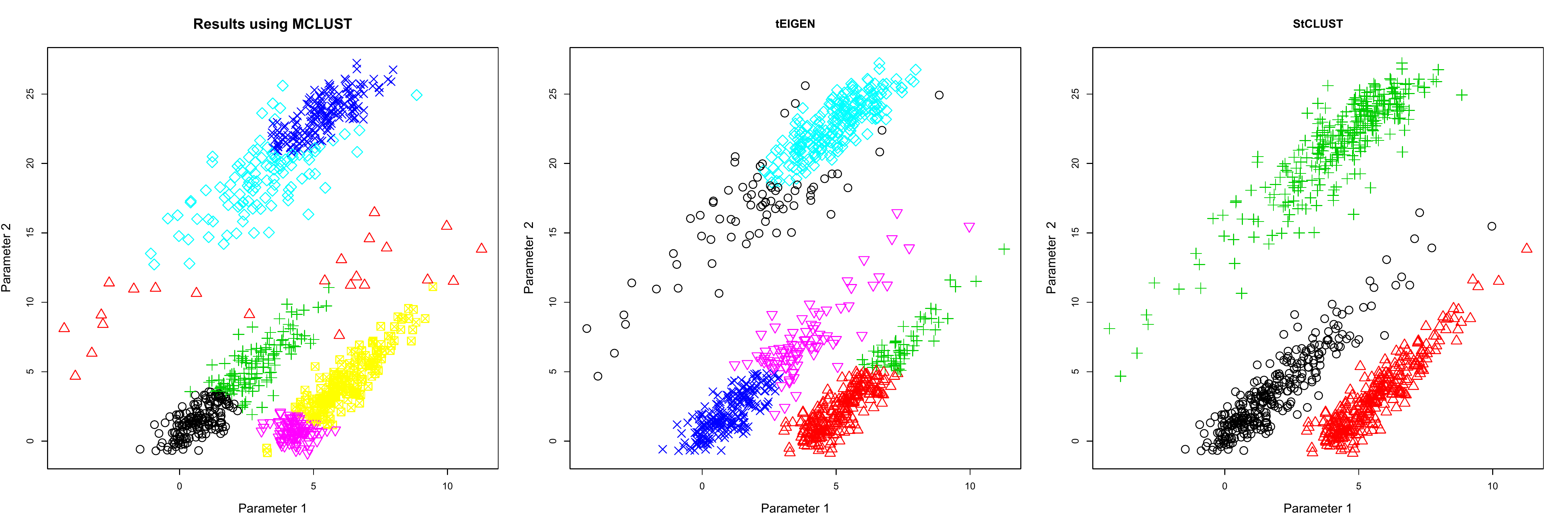}
\caption{The maximum \textit{a posteriori} classification obtained from fitted \MCLUST, \te, and \stclust\ models on one run of simulation~2.}
\label{sim2plot}
\end{center}
\end{figure}

When applied to these simulated data sets, the \stclust\ family of models gave excellent clustering performance, {obtaining an of average ARI of 0.983 and correctly identifying the number of components 90\% of the time (Table~\ref{sim2mle}).  \MCLUST\ and \te, on the other hand, consistently overestimated the number of components and produced inferior ARI scores.} The poor classification performance of the \MCLUST\ and \te\ mixtures can sometimes be mitigated by merging components. For example, if we look at the results from a single run in Figure~\ref{sim2plot}, we see that the classification performance of \te\ can be made equal to that of \stclust\ by merging components. However, merging the \MCLUST\ components here will not make its classification performance equal to that of \stclust. The issue of merging is further considered in Sections~\ref{sec:merge} and~\ref{other14}.
\begin{table}[ht]
\caption{The number of components selected by \MCLUST, \te, and \stclust\ for simulation~2, along with average ARI values and standard deviations.}
\label{sim2mle}
{\scriptsize\begin{tabular*}{1.0\textwidth}{@{\extracolsep{\fill}}cccccccc}
\toprule
\textbf{Groups Selected} & \MCLUST\ &\te\  &\stclust\    \\
\midrule 
3&   & 18\%  &   90\% \\
4&  18\% & 32\%&   6\% \\
5&  18\% & 18\% & 4\%  \\ 
6&   26\%& 20\% &   \\
7&   30\%& 12\% &   \\
8&   2\%&   &   \\
9&   4\%&   &   \\
10&   2\%&   &   \\
\midrule
 Mean ARI  & 0.747  & 0.749  &0.983 \\
Std.\ Dev.\ ARI &0.116  &0.119 & 0.040\\
\bottomrule
\end{tabular*}}
\end{table}

\vspace{+0.6in}

\subsection{Simulation 3}
\label{sim3}
\begin{figure}[htbp]
\vspace{-0.85in}
\begin{center}
\includegraphics[width=0.6\linewidth]{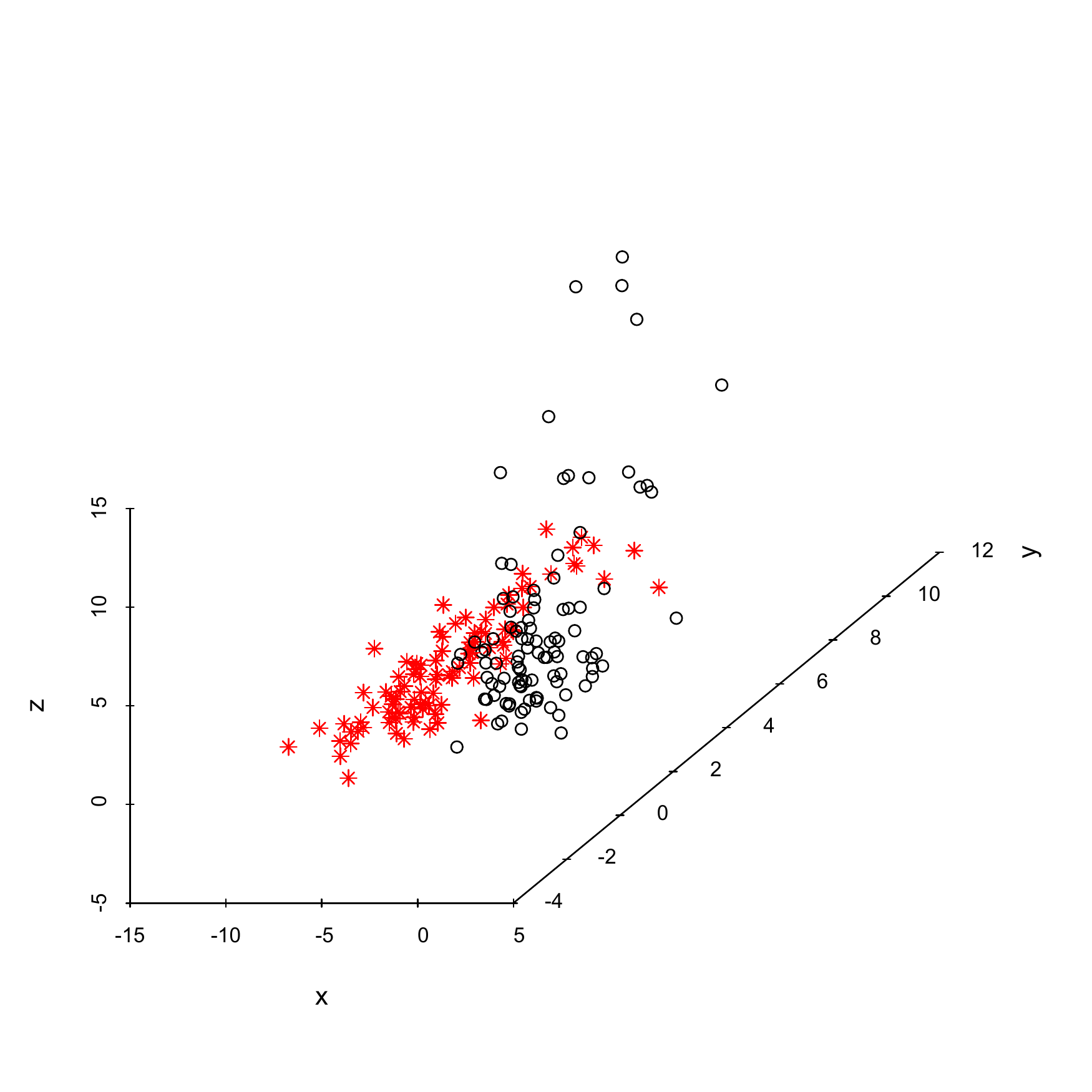}
\vspace{-0.2in}
\caption{{A typical 3D scatterplot of the simulation as described in Section \ref{sim3}, where $\circ$ represents an observation from simulated component~1 and \textcolor{red}{*} represents an observation from a simulated component~2.}}
\label{simSec4plot}
\end{center}
\end{figure}

\begin{figure}[h]
\vspace{-0.6in}
\begin{center}
$\begin{array}{cc}
\includegraphics[width=2.65in]{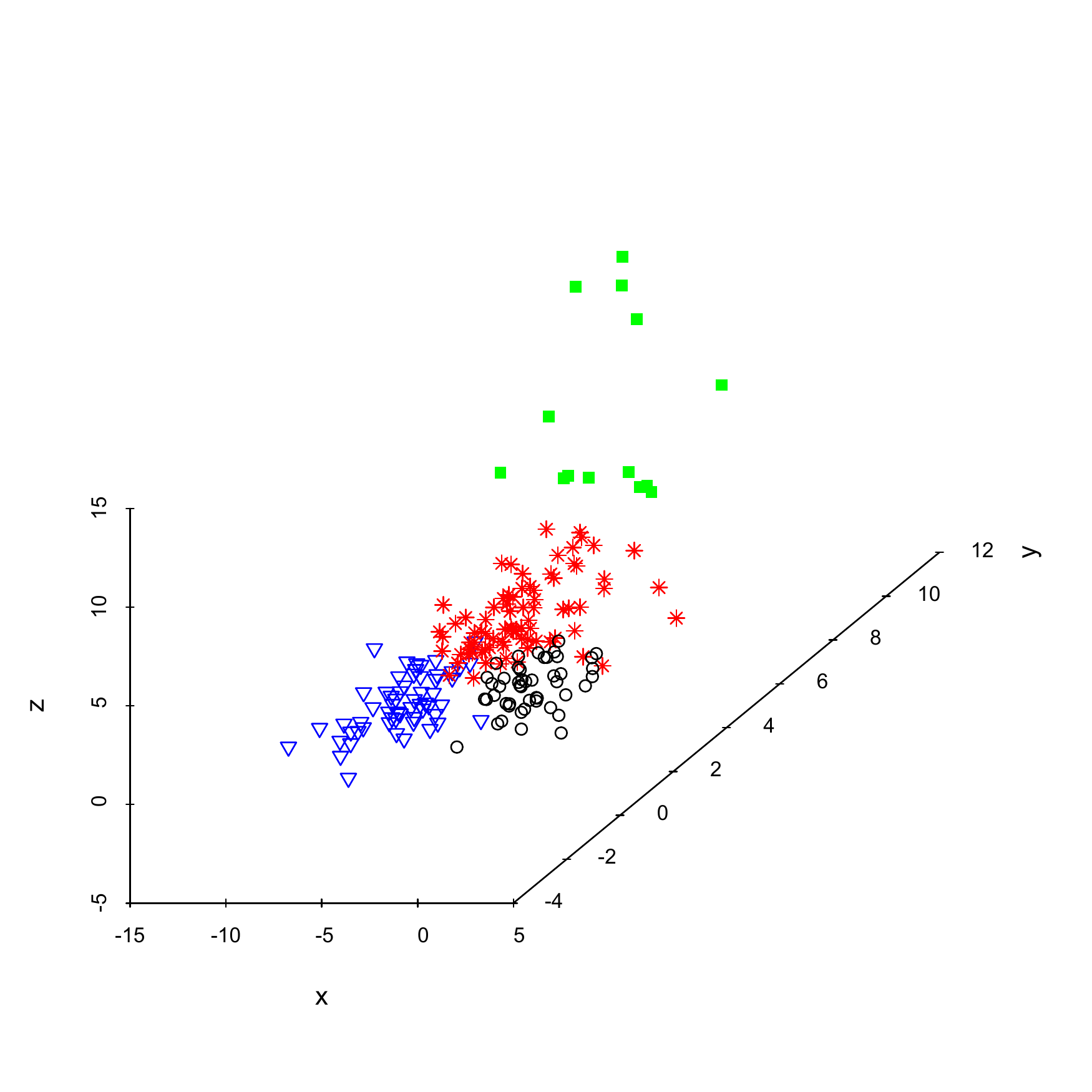} &
\includegraphics[width=2.65in]{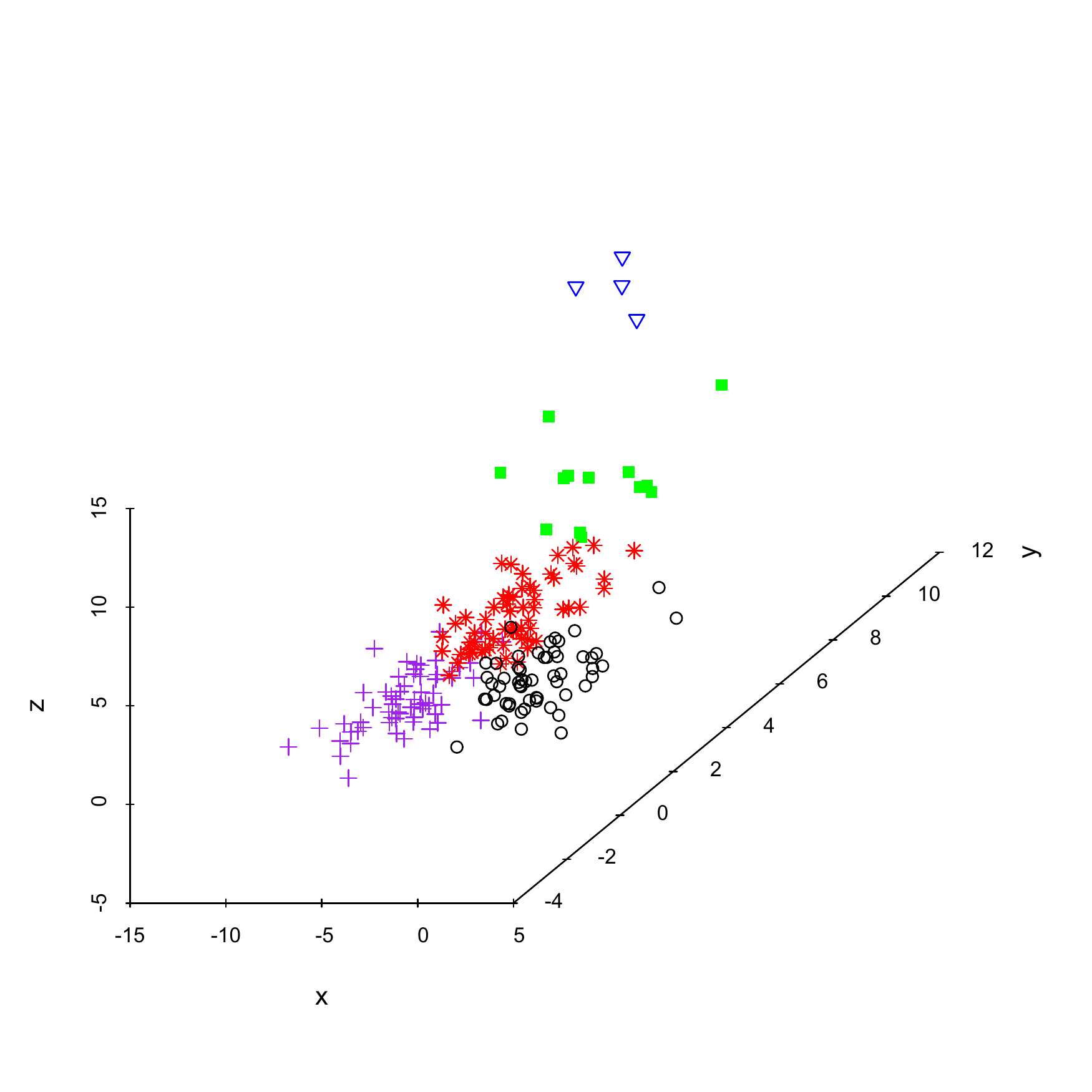} \\[-46pt]
\includegraphics[width=2.65in]{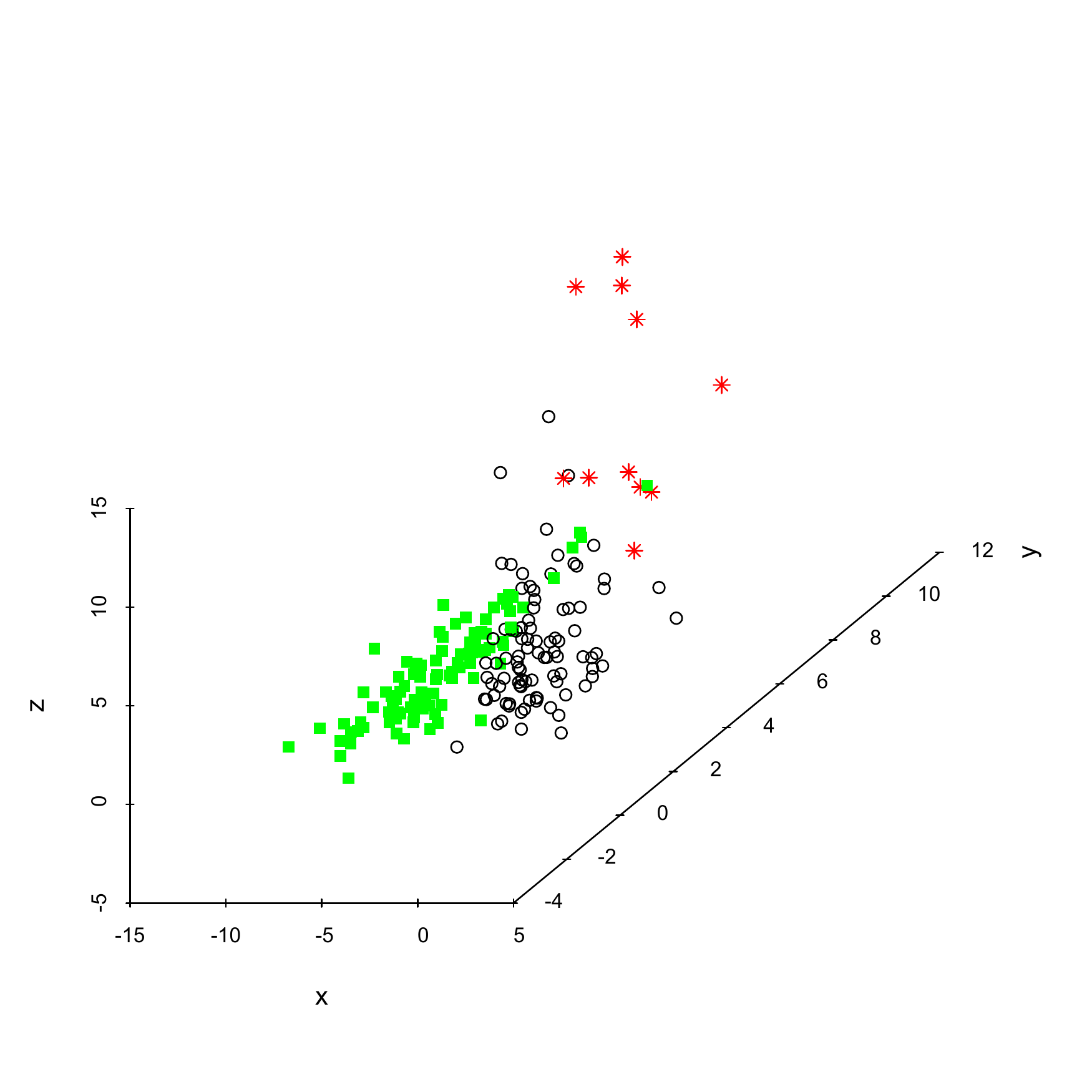} &
\includegraphics[width=2.65in]{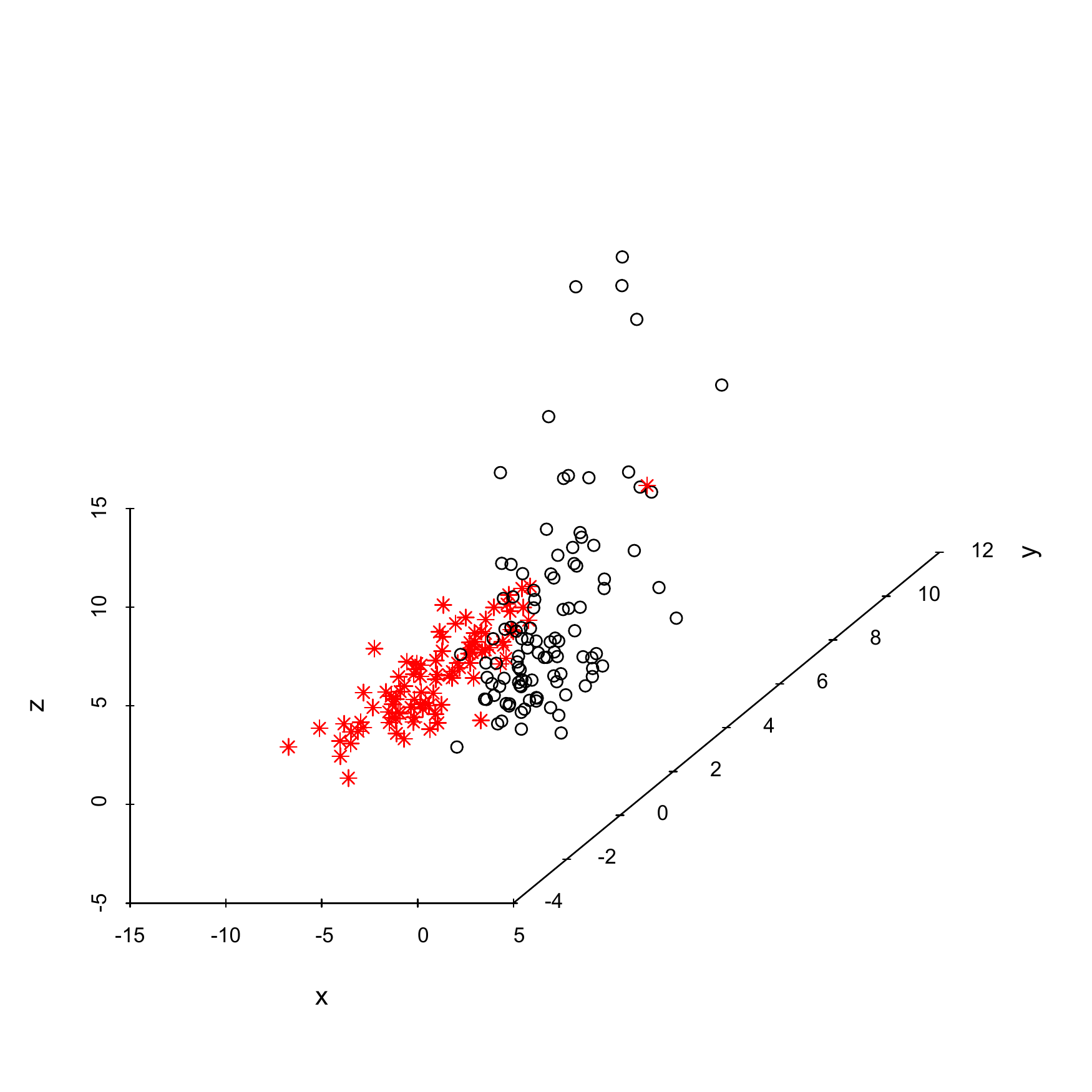}
\end{array}$
\end{center}
\vspace{-0.3in}
\caption{The solutions found by \MCLUST, \te, \snclust, and \stclust\ (top left, top right, bottom left, bottom right, respectively) when applied to the simulated data presented in Figure \ref{simSec4plot}.}
\label{foursome}
\end{figure}

Both simulations in the previous sections represented well separated groups in two dimensions. To test these methods for a more complicated data set, a final simulation study was conducted using two overlapping groups in three dimensions. A two component multivariate skew-$t$ data set was simulated with 100 observations in each group. Typically, the overlapping groups created an X-shaped figure (e.g., Figure \ref{simSec4plot}). One-hundred such random data sets were simulated and the results (Table~\ref{simSec4}) show that the \snclust\ and \stclust\ models perform at least as well as the \MCLUST\ and \te\ families.  Notably, \stclust\ selected a two-component model for 98\% of the simulated data sets {and obtained the highest average ARI}.  Figure \ref{foursome} plots the clustering results corresponding to simulated data in Figure \ref{simSec4plot}.
\begin{table}[ht]
\caption{The number of components selected by \MCLUST, \te, and \stclust\ for simulation~3, along with average ARI values and standard deviations.}
\label{simSec4}
{\scriptsize\begin{tabular*}{1.0\textwidth}{@{\extracolsep{\fill}}cccccccc}
\toprule
\textbf{Groups Selected} & \MCLUST\ &\te\  &\snclust\ &\stclust\    \\
\midrule
2&   6\%& 64\%  &  60\% & 98\%\\
3&   51\%& 12\%  &   38\% & 2\%\\
4&  35\% & 16\%& 2\%  & \\
5&  8\% &  7\%&   &\\ 
6&&1\%&&&\\
\midrule
 Mean ARI  & 0.482  & 0.572  & 0.606 &0.650  \\
 Std.\ Dev.\ ARI &0.106  &0.134 & 0.090 & 0.073\\
\bottomrule
\end{tabular*}}
\end{table}

\section{Applications}
\label{applications}
\subsection{Data Sets}
The efficacy of the \stclust\ and \snclust\ models is demonstrated on four real data sets. The \texttt{bank}, \texttt{crabs}, \texttt{iris}, and \texttt{wine} data are frequently used benchmark data sets for testing the performance of various model-based clustering algorithms. As in Section~\ref{simulation}, we facilitate direct comparison with \MCLUST\ by restricting the \te, \snclust, and \stclust\ families to analogues of the ten \MCLUST\ models for all analyses in Section~\ref{applications}.

The Swiss Bank data \citep{bank} records six different measurements on 100 genuine and 100 counterfeit Swiss banknotes.
The crabs data \citep{campbell} contain the morphological measurements of frontal lobe size, rear width, carapace length, carapace width, and body depth.  The measurements are taken in millimetres on 200 crabs, 50 from each combination of sex (male, female) and colour (blue, orange).
The iris data \citep{anderson,fisher} contain the length and width in centimetres of the sepal and petal of three species of irises.  The data set comprises the measurements of 150 flowers, 50 from each of the various species: \textit{setosa, versicolor}, and \textit{virginica}.  
The wine data \citep{wine} contain 13 chemical and physical properties of wine. The data set comprises measurements from 178 wines from three different cultivars: Barolo, Barbera, and Grignolino.

\subsection{Results}
\label{prevresults}
The \MCLUST, \te, \snclust, and \stclust\ families --- with the latter three restricted to the \MCLUST\ models --- were fitted to the the scaled data for ${i}=1,\dots,9$. Again, in the event that a $g=9$ component model was selected, models were rerun for $i=1,\dots,12$. The chosen models (in accordance with the BIC) are presented with their corresponding ARIs in Table~\ref{results1}; at least one of our two skewed families outperforms both \MCLUST\ and \te\ in each case.
\begin{table}[ht]
\caption{The ARI and members selected for each family when run on the bank, crabs, iris, and wine data sets. For each data set, the best ARI is highlighted in bold face.}
\scriptsize{\begin{tabular*}{1.0\textwidth}{@{\extracolsep{\fill}}lcccc}
\toprule
\hline
& \multicolumn{4}{c}{\textbf{Model Selected, Number of Components, ARI}} \\
 \cmidrule(r){2-5}
 & \MCLUST\ &\te\ &\snclust & \stclust\ \\
 \midrule
Bank& EEE 4, 0.679 & CCCC, 4, 0.679 & EEE, 4, 0.606 & EEE, 2, \textbf{0.980}\\
Crabs & EEE, 4, 0.311 & CCCC, 10, 0.508 & EEE, 4, \textbf{0.838}& EEE, 5, 0.741\\
Iris &VVV, 2, 0.568 & UUUC, 2, 0.568& VEI, 3, \textbf{0.941}&VEI, 3, 0.922\\
Wine&VEI, 8, 0.481 & CICC, 7, 0.454& VEI, 4, 0.801 & EVI, 3, \textbf{0.947}\\
\hline
\bottomrule
\end{tabular*}}
\label{results1}
\end{table}

Focusing first on the \texttt{bank} data set, we see comparable results between the \MCLUST, \te, and \snclust\ families.  However, a great improvement in ARI is achieved when the \stclust\ family is used. The superiority of \stclust\ over the other three families is largely attributable to the fact that it selects the same number of components as true classes. 
The remaining families select a four-component model and so their performance might be improved via merging components (Section~\ref{sec:merge}).

The famous \texttt{crabs} data set is typically classified by sex and species (blue or orange), corresponding to four groups. Although the \MCLUST\ family correctly selects a four-component model, the group memberships do not agree with this known partition  and produced a low ARI (0.311).  The \te\ family obtained a higher ARI (0.508); however, it greatly overestimates the number of groups.   \snclust, which selects a four-component model, performs much better than its competitors. This is not surprising because \stclust\ is indicating skew-normal components ($\hat{\nu}_1 = 90$, $\hat\nu_2 = 89.807$, 
$\hat\nu_3 = 90$, $\hat\nu_4 =  8.455$, $\hat\nu_5 = 90$, $\hat\nu_6 = 90$) and the parameter estimates suggest that the groups are negatively skewed; 
$\hat{\skw}_1 = (-1.4, -1.2, -1.5, -1.6, -1.4)'$, 
 $\hat{\skw}_2 = (-1.9, -1.5, -2.0, -2.0, -1.9)'$,  
$\hat{\skw}_3 = (-1.3, -1.7, -1.4, -1.5, -1.3)'$, \ 
$\hat{\skw}_4 = (-1.3, -1.3, -1.2, -1.2, -1.2)'$.

For the \texttt{iris} data, the ARI obtained for both the \MCLUST\ models and their $t$-analogues are comparable.  Thus, little gain is achieved through the inclusion of the degrees of freedom parameter present in the \te\ and \stclust\ models.  The failure of the addition of the degrees of freedom parameter to improve on the clustering results here might be connected to its limitations as a one-dimensional parameter. Had we used a $p$-dimensional degrees of freedom parameter, allowing different degrees of freedom in each dimension, some improvement may have been observed. This extension will be explored in future work.

The results for the {\tt wine} data set tell a similar story to that of the banknotes. As can be seen in Table~{\ref{results1}, \MCLUST\ and \te\ produce almost identical clustering results; however, a considerable improvement in ARI is apparent once skewness is introduced. As mentioned previously, the increased ARI is greatly owing to the fact that \stclust\ selects the value of $g$ corresponding to the number of true classes.

\subsection{Comparing with Merged Clusters}
\label{sec:merge}
The recent work of \cite{baudry2010combining} argues that the number of mixture components does not necessarily correspond to the number of true groups or clusters.
For instances where an underlying group is comprised of a mixture of two or more Gaussian distributions, the foregoing paper proposes a method for combining mixture components to represent one cluster.  This method first fits a mixture of Gaussian distributions with $g$ components then successively merges mixture components together, resulting in a $k$ component solution where $k \le g$.  The recent version of {\tt mclust} \citep[i.e., version~4.0,][]{fraley2012mclust} contains a function called {\tt clustCombi} that implements the methodology proposed in \cite{baudry2010combining}.  This section compares the clustering results produced by {\tt clustCombi} with the results presented in Section~\ref{prevresults}.

To compare our results with the optimal solution derived by merging components, we used two different merging techniques. The first involves the {\tt clustCombi} function, which starts with the original $g$ component solution (as chosen by the BIC) and combines two components according to an entropy criterion to obtain a $g-1$ component solution.  This procedure is carried out until the one component solution is obtained.  The ARI is calculated at each merging step and the solution with the largest ARI is saved as the `best'. Hereafter we refer this result as the maximum {\tt clustcombi} solution or MCC for short.  The second method involves merging components by hand to create the most advantageous solution.  When looking at a cross tabulation of the known group labels (or true clusters) against the clustering results obtained from the clustering algorithm, this becomes a matter of combining the columns that will maximize the agreements between group labels and components.  An example of this merging process using cross-tabulations (or so-called classification tables) is demonstrated using the \MCLUST\ solution for the {\tt bank} dataset in Table~\ref{gtab}.  Herein, we shall refer to this second merging method as the merging by hand or MBH solution. Note that both the MCC and MBH procedures rely on knowing the true group membership labels;  therefore, they could not be used for real clustering applications.
\begin{table}[h!]
        \caption{Left: Classification results found by \MCLUST\ cross-tabulated against the true group labels. Right: The merging by hand (MBH) classification results (created by merging the second with the third components found by \MCLUST\ and similarly merging the first with the fourth) cross-tabulated against the true group labels.}
        \label{gtab}
        \subtable
            {
\scriptsize{\begin{tabular*}{0.41\textwidth}{@{\extracolsep{\fill}}lcccc}
\toprule
&\multicolumn{4}{c}{\MCLUST} \\
\cmidrule(r){2-5}
 True  & 1 & 2 &3  & 4 \\
\midrule
1&1&75&24&0\\
2&15&0&0&85\\
\bottomrule
\end{tabular*}}\centering
            }
        \subtable
            {
\scriptsize{\begin{tabular}{c}
\\
merge \MCLUST\ \\
components 2 and 3\\
\& components 1 and 4\\
$\rightarrow$
\end{tabular}}\centering
            }
        \subtable
            {
\scriptsize{\begin{tabular*}{0.3\textwidth}{@{\extracolsep{\fill}}lcc}
\toprule
&\multicolumn{2}{c}{MBH} \\
\cmidrule(r){2-3}
True  & 1 & 2 \\
\midrule
1&1&99\\
2&100&0\\
\bottomrule
\end{tabular*}}\centering
            }
\end{table}

Table~\ref{ARIredo} summarizes the ARI values for the pertinent models (indicated using gray font) after performing the MCC and MBH methods described above.  Note that if the true number of clusters is greater than or equal to the number of components found by the clustering algorithm, no gain in ARI can be obtained by merging. 
For instance, when the clustering algorithms under consideration are applied to the {\tt iris} data set (which has three true clusters) either a two or three component solution is obtained and no combination of component merging will result in a better classification. Because no merging is done, the results are unchanged from~Table \ref{results1} and are given in black font in Table~\ref{ARIredo}. Even when a merging technique is used, the \snclust\ and \stclust\ models perform as well as or better than the \MCLUST\ and \te\ families in each case.
\begin{table}[ht]
\caption{ARI values for each data set after merging components is considered. The ARI values calculated after merging components (cf.\ Table~\ref{results1}) are given in grey font. Where merging cannot improve classification performance, the (unchanged) ARI values are given in black font. For each ARI value, the associated number of components is given in brackets and for each data set, the best ARI value is typed in bold face.}
\label{ARIredo}
\scriptsize{
\begin{tabular*}{1.0\textwidth}{@{\extracolsep{\fill}}lrrrrr}
\toprule
\hline
& \multicolumn{5}{c}{\textbf{ARI}}  \\
 \cmidrule(r){2-6}
& \multicolumn{1}{c}{\textbf{MCC}} & \multicolumn{4}{c}{\textbf{MBH}} \\
 \cmidrule(r){2-2} \cmidrule(r){3-6}
  & \MCLUST\  & \MCLUST\ &\te\ & \snclust &\stclust\ \\
 \midrule
Bank& \textcolor{gray}{0.860 (3)}  &  \textcolor{gray}{\textbf{0.980} (2)} & \textcolor{gray}{\textbf{0.980} (2)}   & \textcolor{gray}{\textbf{0.980} (2)} & \textbf{0.980} (2)\\
Crabs&  0.311 (4) &  0.679 (4) &  \textcolor{gray}{0.783 (4)}  & \textbf{0.838} (4) & \textcolor{gray}{0.753 (4)}    \\
Iris &0.568 (2)&0.568 (2)& 0.568 (2)& \textbf{0.941} (3) & 0.922  (3)\\
Wine &  \textcolor{gray}{0.876 (5)} &  \textcolor{gray}{0.901 (3)} & \textcolor{gray}{0.867 (3)} & \textcolor{gray}{\textbf{0.947} (3)}& \textbf{0.947} (3)   \\
\hline
\bottomrule
\end{tabular*}}
\end{table}


\section{Model-Based Classification}\label{sec:mbclass}
\subsection{The Model}
In the event that a subset of the data under consideration has known group membership labels, model-based classification can be used.  This is a semi-supervised version of model-based clustering that uses both the labeled and unlabelled data to produce parameter estimates. 
If we order the data such that the first $m$ observations ($m<n$) have known labels, then the Gaussian model-based classification likelihood is
\begin{equation*}\begin{split}
\mathcal{L}(\varthet\mid\vecx_1,\ldots,\vecx_n,\vecz_1,\ldots,\vecz_m) =  \prod_{j=1}^{m}\prod_{i = 1}^g&[\pi_i \phi(\vecx_j\mid\vecmu_i,\matsig_i)]^{z_{ij}}\prod_{k=m+1}^{n}\sum_{l = 1}^h \pi_l \phi(\vecx_k\mid\vecmu_l,\matsig_l),
\end{split}\end{equation*}
%
for $h\geq g$; often, as in our analyses (Sections~\ref{sec:oliveoil}, \ref{classother}, and~\ref{other14}), it is assumed that $h=g$. 

We apply the \snclust\ and \stclust\ families for model-based classification, drawing comparison with the \MCLUST\ and \te\ families. In addition to the real data that we used to illustrate our clustering approaches, we analyze real food authenticity data on Italian olive oils. To facilitate direct comparison with MCLUST, we restrict the \te, \snclust, and \stclust\ families to analogues of the ten models used in \MCLUST\ (cf.\ Table~\ref{ta:steigen}) for the analyses in Sections~\ref{sec:oliveoil} and~\ref{classother}. However, we consider all 14 models (i.e., all 14 constraints for the decomposition of the component scale matrices) for the \snclust\ and \stclust\ families in Section~\ref{other14}.


\subsection{Olive Oil Data}\label{sec:oliveoil}
The olive oil data \citep{forina1982,forina1983} contain the percentage of eight fatty acids found in 572 Italian olive oils.  {The data set are available within the {\tt pgmm} package \citep{mcnicholas11pgmm} for {\sf R}}.  Each oil sample comes from one of three distinct regions, which can be further partitioned into nine areas. \cite{mcnicholas10c} used these data to illustrate model-based classification, by taking random subsets of 50\% of the observations to have known component membership labels; we will do the same to illustrate our models.
 
We first consider the problem of classifying the olive oils into their regions. The four families were run using 15 different random subsets with 50\% of the labels known and $g=3$.  As advocated by \cite{andrews2011}, we employ a uniform initialization whereby the unknown observations have initial predicted labels $\hat{z}_{ij} = 1/g$, for $i=1,\ldots,g$ and $j=1,\ldots,n$.  A similar approach is repeated for the $g=9$ component models, this time classifying the data by geographical area. The average ARI values for the fitted models (Table~\ref{olive}) confirm the slightly superior performance of the \snclust\ family. However, we note that all four families give good classification performance when classifying by region and by area. 

We notice that the \stclust\ models obtain a lower ARI than the \te\ models when classifying the olive oil into geographical area.  One possible explanation is that uniform initialization is not suitable for the \stclust\ family. Because these procedures are susceptible to converge to local maxima, and considering the relatively complicated likelihoods associated with the \skewt\ family, perhaps a more guileful approach such as a deterministic annealing \EM\ \cite[cf.][]{Ueda} is necessary. Further investigation of this idea will be a subject of future work.
\begin{table}[ht]
\caption{The average ARI values, with standard deviations, for the fitted \MCLUST, \te, \snclust, and \skewtCLUST\ models for the olive data by region ($g=3$) and area ($g=9$). For each data set, the best ARI is highlighted in bold face.}
{\scriptsize\begin{tabular*}{1.0\textwidth}{@{\extracolsep{\fill}}lcccccccc}
\toprule
\hline
 & \MCLUST\ &\te\ & \snclust & \stclust\ \\
 \midrule
Olive (Region) & 0.9962 (0.0036) & {0.9965}(0.0032) & \textbf{0.9983} (0.0028) & 0.9946 (0.0021) \\
Olive (Area) & {0.9091} (0.0333) & 0.9185 (0.0224) & \textbf{0.9249} (0.0242) &0.8994 (0.0307) \\
\hline
\bottomrule
\end{tabular*}}
\label{olive}
\end{table}

\subsection{Other Data}
\label{classother}
The same four data sets considered in Section \ref{applications} are used for model-based classification; again, 50\% of the labels are taken to be known. Each data set was run with 25 different random subsets of known labels with the number of groups specified to the correct value. Although the \MCLUST\ and \te\ families benefit substantially under a classification framework, \snclust\ obtained the highest ARI on all four data sets (Table \ref{classres}). Comparing \stclust\ with the its symmetric alternative \te, we see an improvement in ARI for all but the crabs dataset. As mentioned in Section \ref{sec:oliveoil}, this inconsistency could be a result of the uniform initialization.
\begin{table}[ht]
\caption{Average ARI values, with standard deviations, for the fitted \MCLUST, \te, \snclust, and \skewtCLUST\ models. For each data set, the best ARI is highlighted in bold face.}
{\scriptsize\begin{tabular*}{1.0\textwidth}{@{\extracolsep{\fill}}lcccccccc}
\toprule
\hline
 & \MCLUST\ &\te\ & \snclust & \stclust\ \\
 \midrule
Bank & 0.9760 (0.0200) &0.9760 (0.0200) & \textbf{0.9911} (0.0102) &0.9792 (0.0204) \\
Crabs & 0.8798 (0.0474) & {0.8818} (0.0451) &\textbf{0.9255} (0.0389) & 0.8575 (0.0632) \\
Iris &0.8910 (0.0311) & 0.8910 (0.0311) & \textbf{0.9447} (0.0305) & {0.9199} (0.0574) \\
Wine & 0.9004 (0.0396) & {0.8271} (0.0505) & \textbf{0.9444} (0.0504) & 0.8379 (0.0770) \\
\hline
\bottomrule
\end{tabular*}}
\label{classres}
\end{table}

\subsection{The EVV, EVE, VEE and VVE models}
\label{other14}
Heretofore, we restricted the \snclust\ and \stclust\ families to correspond to the ten \MCLUST\ models. This restriction was appropriate to facilitate direct comparison with \MCLUST. In this section, we will consider the full \snclust\ and \stclust\ families; i.e., the \snclust\ and \stclust\ families with all 14 models (Table~\ref{ta:steigen}). We apply the full \snclust\ and \stclust\ families for model-based classification of the data sets considered in Sections~\ref{sec:oliveoil} and~\ref{classother}. We compare the performance of the full families to the versions restricted to correspond to \MCLUST. Note that we implement the four `extra' models for the \snclust\ and \stclust\ families using the majorization-minimization (MM) algorithm \citep[cf.][]{hunter04} developed by \cite{browne13}.

Table \ref{class14} presents ARI scores associated with the reduced (i.e., MCLUST-analogous) as well as the full \snclust\ and \stclust\ families. The last two columns summarize the proportion of runs where one of the four `extra' models was chosen (these correspond to the EVV, EVE, VEE, and VVE models in Table \ref{ta:steigen}).  Our results demonstrate that one of these models is often chosen, resulting in a higher average ARI score for all but one of the \stclust\ runs as well as half of the \snclust\ runs. To be more specific, EVE was selected between 59\% and 100\% of the time for all but the olive oil classification by region, where the EVE model was never selected. We therefore advocate the inclusion of all fourteen models in practice rather than the subset of ten models used by \MCLUST.
\begin{table}[ht]
\caption{Average ARI values, with standard deviations, for the fitted \snclust, and \skewtCLUST\ using the ten model (original) and 14 model (full) families. For each data set, the best ARI is highlighted in bold face.}
{\scriptsize\begin{tabular*}{1.0\textwidth}{@{\extracolsep{\fill}}p{15mm}cccccc}
\toprule
\hline 
& \multicolumn{2}{c}{Original 10 model family} &\multicolumn{2}{c}{Full 14 model family} & \multicolumn{2}{c}{\centering \% of non-\MCLUST}   \\
 \cmidrule(r){2-3}  \cmidrule(r){4-5} \cmidrule(r){6-7}
&\snclust & \stclust\ & \snclust & \stclust\ & \snclust & \stclust \\
 \midrule
Bank &  {0.9911} (0.010) &0.9792 (0.020) & 		0.9908 (0.010)  	&\textbf{0.9896} (0.010) &100\% &100\%	\\
Crabs &\textbf{0.9255} (0.039) & 0.858 (0.063)& 		0.9109 (0.028)  	& 0.9170 (0.033) & 93\% &92\% \\
Iris &{0.9447} (0.031) & {0.9199} (0.057) &	{0.9622} (0.024)		& \textbf{0.9658} (0.020) &98\%& 100\%\\
Wine & {0.9444} (0.050) & 0.8379 (0.077)&	 	0.9374 (0.046)		& \textbf{0.9474} (0.030) & 64\%&85\% \\
Olive(Region) & 0.9981 (0.004) & {0.9965}(0.003) & \textbf{0.9994} (0.002) & {0.9983} (0.002) & 100\% &  67\%\\
Olive(Area) & {0.9091} (0.033) & 0.9185 (0.022) & \textbf{0.9655} (0.012) & {0.9562} (0.011) & 100\% & 100\%\\
\hline
\bottomrule
\end{tabular*}}
\label{class14}
\end{table}

\section{Concluding Remarks}\label{conclusions}

This paper builds on the growing trend towards non-Gaussian model-based clustering by developing two families of models that account for skewness: skew-normal and \skewt\ analogues of the GPCM family of models. Parameter estimation was outlined and our novel families were applied to simulated and real data. Both model-based clustering and classification were illustrated using several real data sets, and the performance of \stclust\ and \snclust\ was generally superior to their symmetric analogues. Interestingly, this superior performance was often retained even after merging components was considered.

Future work will focus on the initialization of these families in addition to the search for more effective model selection techniques. As briefly mentioned in Section~\ref{prevresults}, we could investigate the efficacy of extending the degrees of freedom to be $p$-dimensional. From the classification viewpoint, we only considered the semi-supervised model-based classification scenario, where one mixture component corresponds to a class. The straightforward extension to model-based discriminant analysis will be investigated and incorporated into {\sf R} packages that are being developed for the \stclust\ and \snclust\ families. Finally, the use of our skewed families in the analysis of data of mixed type \citep[cf.][]{browne12b} will also be investigated.

\section*{Acknowledgements}
This work was supported by an Ontario Graduate Scholarship, a Discovery Grant from the Natural Sciences and Engineering Research Council of Canada, and an Early Researcher Award from the Ontario Ministry of Research and Innovation.

\bibliographystyle{chicago}

\end{document}